\documentclass[acmtocl]{acmtrans2m}
\acmVolume{1}
\acmNumber{1}
\acmYear{08}
\acmMonth{May}

\newcommand{\BibTeX}{{\rm B\kern-.05em{\sc i\kern-.025em b}\kern-.08em
    T\kern-.1667em\lower.7ex\hbox{E}\kern-.125emX}}

\title{Confidentiality, Integrity and High Availability with Open Source IT green}
\author{Luciana Guimaraes}

\begin{abstract} 
This paper presents elements that form the structure of a network of data using secure stable and mature technologies that meet the requirement of having code free. The principle would be conflicting code open Tuesday where he wants to keep maximum control over the data but is already evidence that open source does not hide the famous backdoor possible in closed systems code.

Basearemos this work experience gained in a real environment and using paravirtualization to show a situation more critical and now real in most companies, the virtualization of servers.
\end{abstract}

\category{k.6.3}{Management of Computing and Information Systems}{Software Management}%
[software selection]



\begin{document}

\begin{bottomstuff} 
Authors' addresses:  Luciana Guimarães, Docents, University of Managerial Sciences UNA, Brazil-Minas Gerais-Belo horizonte, 30570-310
\end{bottomstuff}

\maketitle

\section{Introduction}

By working in a company which provides service to the network of municipalities my
company is subject to any kind of attack, either via the Internet, social attacks, as in our own Intranet and Extranet by political enemies of our customers.
Seeing this picture began to plan a way to keep information secure as planned and located most critical points in the structure, was necessary to create a map of where each risk manager should define on a scale of zero to ten, on'ıvel criticality that the loss would have a certain appeal, being listed as resources to Phone ˆ onia, the network of data, the computers and printers documents into folders at'e fax equipment.
In this article we point out solutions to all these points without it being necessary spent on purchase of software and more important with the use of technologies already established as stable in their specialties.

\section{METHODOLOGY}

\subsection{PLATFORM OF TESTS}
We'll laboratory tests with the following equipment;
\subitem 2 units with the following characteristics, Cpu dual core 1.6GHz, 1GB RAM, 80 GB HD. They will be our primary and secondary servers.
	
\subitem  2 units with the following characteristics, 1.6 GHz Pentium CPU, 500 MB RAM, 40 GB HD. These units faram role of our estaÃ Ã § μ is the work being with a Windows operating system and another with Linux Debian.
\subitem 1 Switch  8/100 Mb/s

\subsection{THE ARCHITECTURE OF THE PLATFORM OF TESTS }

\subsection{POSTGRESQL}

PostgreSQL is a powerful, relational database system open source. It has more than 15 years of activity and development of this architecture has a strong reputation for reliability, data integrity and accuracy. It runs on all major operating systems, including Linux, UNIX (AIX, BSD, HP-UX, SGI IRIX, Mac OS X, Solaris, Tru64), and Windows. is fully compatible ACID has the full support of foreign keys, joins, views, triggers and stored procedures (in go ¡rivers languages). It includes more types of data SQL92 and SQL99, including INTEGER, NUMERIC, BOOLEAN, CHAR, VARCHAR, DATE, INTERVAL, and TIMESTAMP. It also supports storage of binary large objects, including images, sounds or video. It supports native programming interfaces for C / C + +, Java,. Net, Perl, Python, Ruby, Tcl, ODBC, among others, and exceptional documentation.
Why does not the Postgresql and Mysql? Optei for using Postgresql because it totally free and no matter the platform to be used. He has control of transactions is more mature and more stable and easier to restore in case of panes in hardware.

\subsection{SAMBA}
	
Samba is a service, used in UNIX-type operating systems, which simulates a Windows server, enabling management to be done and file sharing in a Microsoft network.
In version 3, Samba are not files and provides printing services to various clients Windows, but can also integrate itself with Windows Server Domain, both as a Primary Domain Controller (PDC) or as a Domain Member. It may also be part of an Active Directory Domain.
From recognized stability in the process of linking different platforms. In will have an environment with Windows and Linux machines working only with a source of files / data.

\subsection{NFS} 
	
NFS - File System Network (Network File System). Protocol used to access the file systems on a network. It is possible to mount file systems of other machines through this protocol.
The NFS is faster and more stable than the SAMBA but does not allow the interconnection between Windows and Linux without the need for the purchase of a software client / server to the side windos therefore only be used to interconnect machines with Linux.

\subsection{OPENSSH}

OpenSSH is a free version of the tools of connectivity SSH uses techniques that users of the Internet can trust. OpenSSH encrypting all traffic (including passwords) to effectively eliminate the eavesdropping, connection of kidnappings and other attacks. Moreover, provides OpenSSH tunneling and various methods of authentication, and supports all versions of SSH protocol.
In case of connection between equipment and will need to prompt or graphical environment we using SSH, SCP for the transfer of data over the network is encrypted.
\subsection{HEARTBEAT}
	
The project Linux-HA (High-Availability Linux) focuses on research and implementation of solutions for high availability (clustering) for Linux.
The main component of this project in development is the heartbeat that works as manager of the cluster and its resources. As the name indicates, signalling the presence (or absence) of contact with the nodes of the cluster is made by sending heartbeats of small packages addressed to all nodes in the cluster, whose confirmation of receipt by each node indicates the state that node.
This product enters the model as a guardian of servers tracking any service that is necessary. In our article we are monitoring the services of the database, ssh, ssl, http, https.

\subsection{DRBD }
	
DRDB is a device designed to build blocks of clusters of high availability. This is done by mirroring a whole block of the device via the network. It will be responsible for the replication of each bit stored in the server's main winchester

\subsection{APACHE2}
	
The Apache HTTP Project is a collaborative effort to develop software that aims to create the implementation of an HTTP server (Web) and solid open source. The project is managed jointly by a group of volunteers located around the world, using the Internet and the Web to communicate, plan and develop the server and its documentation. This project is part of the Apache Software Foundation. In addition, hundreds of users contribute ideas, code and documentation for the project.
As more robust the security point of view we are using this version.

\subsection{PHP5}
	
PHP (a recursive acronym for "PHP: Hypertext Preprocessor) is a programming language for computers interpreted, free and widely used to generate dynamic content on the web. Despite being a language of learning and easy to use for small dynamic simple scripts, PHP is a powerful oriented language the objects.
Despite being new we are using this tool as the PHP4 is not the object oriented and is no longer being held and that is complicating factor because we need to be not only to this but with all packages always updated with regard to the question less security .

\subsection{SNORT}
	
Snort is a free software to prevent invasions of the Network (NIPS) and intrusion detection network (NIDS) capable of carrying out analysis of traffic in real time over IP networks.
Snort runs of protocol analysis, content searching, and is commonly used to actively or passively block a variety of attacks and crawls, such as buffer overflows, stealth port scans, attacks on aplicaçõe web, tracking the SMB, and attempts to simulation of SO , Among other characteristics. The software is used mainly for prevention of intrusion, Snort can be combined with other software, as SnortSnarf, sguil, OSSIM, and the Basic Analysis and Security Engine (BASE) to provide a visual representation of intrusion. With patches for the Snort it offers support for packet stream and virus scanning as ClamAV and with the SPADE abnormalities in the network can be found in layers 3 and 4 through analize history.

\subsection{IPTABLE}
	
It will be responsible for the blocking of services, machines and packages that are not allowed to travel on the network.

\subsection{XEN}
	
The Xen hypervisor that provides a powerful, efficient and safe for use virtualization for x86 CPUs, x8664, IA64, PowerPC and other architectures, has been used to virtualize a wide range of clients and operating systems, including Windows, Linux, Solaris and several versions of the BSD operating systems. It is widely regarded as an attractive alternative to proprietary platforms and virtualization hypervisors for x86 platforms and IA64.

\subsection{EXT3}
	
The ext3 (which means "third extended file system") is part of the new generation of management systems, the Linux file. Its biggest advantage is the support of journaling, which is to store information on the transactions of writing, allowing a rapid and reliable recovery in case of sudden interruption (for example, for lack of electricity).
Use of this file system improves the recovery of the file system in case of any sudden shutdown of the computer, through sequential recording of data in the area of metadata and access mhash of its directory tree

\section{RESULTS OF TESTS}
\subsection{STRATEGY OF TESTS}
	
We set up the equipment as shown in the following sections and after that start the testing process and cominucação using micro-specific benchmarks for this purpose.
We chose a database and an application Postgresql testarmos PHP for the fall issue of reactivation of the equipment and checking time to return to normal operations, the rate of transfer to upgrade the base replicated, time of activation of mirror machine.
Below enumeramos the methodologies used for testing of tolerance is divided into two parts and using disks or system failure in LVM, one of the machines failed the physical hardware and one of the servers:

\subitem Part 1 - PHP processing.

\subitem Part 2 - Processing of the bank Postgresql.

\subitem failed Server 1
\subsubitem Server 1 is running the virtual machines vm1 and vm2
\subsubitem Server 2 is the virtual machines running vm3 and vm4
\subsubitem Server 1 is off or has defects in operation
\subsubitem Heartbeat in Server 2 detects failure of the Server 1
\subsubitem Heartbeat boots virtual machines vm1 and vm2 in Server 2
\subsubitem Server 1 is restored
\subsubitem Heartbeat in Server 1 if communicates with a Heartbeat Server 2
\subsubitem Heartbeat in Server 2 paralyzes the virtual machines vm1 and vm2
\subsubitem Heartbeat in 1 Server virtual machines vm1 boots and vm2
\subsubitem service returns to normal

\subitem failed Server 2
\subsubitem Server 2 has vm1 virtual machines and vm2
\subsubitem Server 1 has subsubitem Server virtual machines vm3 and vm4
\subsubitem Server 2 is switched off or has defects in operation
\subsubitem Heartbeat in Server 1 detects failure of Server 2
\subsubitem Heartbeat boots virtual machines vm3 and vm4 in Server 1
\subsubitem Server 2 is restored
\subsubitem Heartbeat in Server 2 would communicate with Heartbeat in Server 1
\subsubitem Heartbeat in Server 1 paralyzes the virtual machines vm3 and vm4
\subsubitem Heartbeat Server 2 boots in virtual machines vm3 and vm4
\subsubitem service returns to normal

These tests were failures of tolerance will be made as follows:
Simulation of the failure of the server by stopping the service of heartbeat

Simulating the failure of the server 1, enter the following command in the server 1:
/etc/init.d/heartbeat stop

Stop the server through its forced shutdown (pulling power cord from the)
Stop the server through its disengagement correct. (command 'shutdown')

\subsection{MOUNTING CONFIGURATION AND THE ENVIRONMENT}
\subsubsection*{Installation of the Linux operating system Debian Etch}
	
On devices defined as servers. We will be using this distribution by the stable version available on the date of creation of this article.
In this installation use partitioning EXT3 for installation of data, and the division of HD in our area of 2.7 GB to SWAP and the rest of the unit for data.

\subsubsection*{Instalation NFS}
	
sudo aptitude install nfs-common nfs-server-kernel portmap

Once installed the packages edit /etc/exports and add the directories to be accessed remotely, see the example below:

/u/usr 10.0.2.6 (rw, sync)

Above are sharing the directory /u/ usr only to the machine 10.0.2.6 allowing this writing and reading and forcing syncronismo between the two machines.

\subsubsection*{Installing SAMBA}

sudo aptitude install smbfs samba samba-common smbclient

Edit /etc/samba/smb.conf and observe the following parameters:

workgroup = XXXXXXXX

server string = XXXXXXXX

printcap name = /etc/printcap

load printers = no

socket options = TCP\_NODELAY SO\_RCVBUF=8192 SO\_SNDBUF=8192

dns proxy = no

netbios name = padrao

netbios aliases = padrao

map to guest = never

os level = 99

preferred master = no

domain master = no

wins support = no

dead time = 0

domain logons = no

printcap name = cups

printing = cups

log file = /var/log/samba/log.\%m

max log size = 50

debug level = 1

security = share

unix password sync = yes

password level = 0

null passwords = yes

encrypt passwords = true

smb passwd file = /etc/samba/smbpasswd

username map = /etc/samba/smbusers

username level = 8

add machine script = /usr/sbin/adduser -n -r -g machines -c "Samba machine" -d /dev/null -s /bin/false \%u

passdb backend = smbpasswd

idmap uid = 16777216-33554431

idmap gid = 16777216-33554431

template shell = /bin/false

winbind use default domain = no

bind interfaces only = no

hide dot files = no

[Desenv]

comment = XXXXXXXXXXXXXXXx

path = /XXXXXXXXXX

public = no

browseable = yes

guest only = no

guest ok = yes

writable = yes

preserve case = No

short preserve case = No

directory mask = 0777

valid users = luciana

create mask = 0777

available = yes

\subsubsection*{Installing DRDB}

The advantage of DRDB8 on SRDB7 are: It allows resources to be master of both the time and can be mounted with Permissions of reading and writing. Now we will compile the modules from DRDB8 to be loaded into the kernel. For this we need the packages â build-essential and kernel-headers-xen. Do intão the prompt;

sudo aptitude install drbd8-utils drbd8-module-source drbd8-source build-essential linux-headers-xen
sudo sudo m-a-i-module drbd8-source
sudo update-modules
sudo modprobe drbd

This will compile the modules for kernel / drivers / block / drbd.ko and will be used for this kernel. A configuration padão was set up in / etc / drbd.conf

Configuration:

Edit o /etc/drbd.conf

global {
    usage-count yes;
}

common {
  syncer { rate 10M; }
}

resource r0 {
  protocol C;
  handlers {
    pri-on-incon-degr "echo o > /proc/sysrq-trigger ; halt -f";
    pri-lost-after-sb "echo o > /proc/sysrq-trigger ; halt -f";
    local-io-error "echo o > /proc/sysrq-trigger ; halt -f";
    outdate-peer "/usr/sbin/drbd-peer-outdater";
  }

  startup {
  }

  disk {
    on-io-error   detach;
  }

  net {
    allow-two-primaries;
    after-sb-0pri disconnect;
    after-sb-1pri disconnect;
    after-sb-2pri disconnect;
    rr-conflict disconnect;
  }

  syncer {
    rate 10M;
    al-extents 257;
  }

  on node1 {
    device     /dev/drbd0;
    disk       /dev/sda3;
    address    192.168.0.128:7788;
    flexible-meta-disk  internal;
  }

  on node2 {
    device    /dev/drbd0;
    disk      /dev/sda3;
    address   192.168.0.129:7788;
    meta-disk internal;
  }
}

"Allow-two-primaries" option that allows you to be mounted as master "master" at the beginning of our network. Copy /etc/drbd.conf for o node 2 and restart drbd with the following command.
sudo / init.d / drbd restart

If you want to check the state run the command below

sudo /etc/init.d/drbd status

This should be the response if everything is OK.

drbd driver loaded OK; device status:

version: 8.0.3 (api:86/proto:86)

SVN Revision: 2881 build by root@node1, 2008-01-20 12:48:36
 0: cs:Connected st:Secondary/Secondary ds:UpToDate/UpToDate C r---
    ns:143004 nr:0 dw:0 dr:143004 al:0 bm:43 lo:0 pe:0 ua:0 ap:0
        resync: used:0/31 hits:8916 misses:22 starving:0 dirty:0 changed:22
        act\_log: used:0/257 hits:0 misses:0 starving:0 dirty:0 changed:0

replace the appeal to the master with the following command in equipment

sudo drbdadm primary r0

and check the status again

sudo /etc/init.d/drbd status

drbd driver loaded OK; device status:

version: 8.0.3 (api:86/proto:86)

SVN Revision: 2881 build by root@node1, 2008-01-20 12:48:36
 0: cs:Connected st:Primary/Primary ds:UpToDate/UpToDate C r---
    ns:143004 nr:0 dw:0 dr:143004 al:0 bm:43 lo:0 pe:0 ua:0 ap:0
        resync: used:0/31 hits:8916 misses:22 starving:0 dirty:0 changed:22
        act\_log: used:0/257 hits:0 misses:0 starving:0 dirty:0 changed:0

As you can see action is "master" in both of us device. And the drbd is now accessible on / dev/drbd0

File system
	
We can now create the file system in / dev/drbd0 with the following command

sudo mkfs.ocfs2 /dev/drbd0

This can be mounted simultaneously in both with the commands below:

sudo mkdir /drbd0

sudo mount.ocfs2 /dev/drbd0 /drbd0

Now we have a syncronismo between storage devices.

Init script

We have to make sure that, after restart, the system will set drbd resources again to "master" and mount a "/ drbd0" before starting the Heartbeat and Xen machines.

Edit /etc/init.d/mountdrbd.sh

drbdadm primary r0 

mount.ocfs2 /dev/drbd0 /mnt
 	
make a symbolic link to executable / etc/rc3.d/S99mountdrbd.sh

sudo chmode +x /etc/init.d/mountdrbd.sh

sudo ln -s /etc/init.d/mountdrbd.sh /etc/rc3.d/S99mountdrbd.sh

In fact, this step can also be integrated to Heartbeat, adding adequate resources for the setting. But as time is that vai do with this script.

\subsubsection*{Installation Heartbeat2}
	
Now we can install and configure the Heartbeat 2

sudo apt-get install heartbeat-2 heartbeat-2-gui

Edit /etc/ha.d/ha.cf

crm on

bcast eth0

node node1 node2

restart heartbeat2 com

sudo /etc/init.d/heartbeat restart

\subsection{Startup}

Edit the file /root/cluster/bootstrap.xml

 cluster\_property\_set id="bootstrap"

attributes

  nvpair id="bootstrap01" name="transition-idle-timeout" value="60"/

  nvpair id="bootstrap02" name="default-resource-stickiness" value="INFINITY"/

  nvpair id="bootstrap03" name="default-resource-failure-stickiness" value="-500"/

  nvpair id="bootstrap04" name="stonith-enabled" value="true"/

  nvpair id="bootstrap05" name="stonith-action" value="reboot"/

  nvpair id="bootstrap06" name="symmetric-cluster" value="true"/

  nvpair id="bootstrap07" name="no-quorum-policy" value="stop"/

  nvpair id="bootstrap08" name="stop-orphan-resources" value="true"/

  nvpair id="bootstrap09" name="stop-orphan-actions" value="true"/

  nvpair id="bootstrap10" name="is-managed-default" value="true"/

 /attributes

/cluster\_property\_set

Load the file with the following command

sudo cibadmin -C crm\_config -x /root/cluster/bootstrap.xml

This will start the Cluster with the values set in xml file

Setting up the device STONITH

Using the command below the keys to create trust between the servers.

sudo ssh-keygen

-- save key under /root/.ssh/*

-- dont give any passphrase

scp /root/.ssh/id\_rsa.pub node2:/root/.ssh/authorized\_keys
	
Now make sure you can log on the server 2 from the server 1 without using password.

sudo ssh -q -x -n -l root "node2" "ls -la"

Stonith of configuring the server 2

/root/cluster/stonith.xml

clone id="stonithclone" globally\_unique="false"

 instance\_attributes id="stonithclone"

  attributes

   nvpair id="stonithclone01" name="clone\_node\_max" value="1"/

  /attributes

 instance\_attributes

 primitive id="stonithclone" class="stonith" type="external/ssh" provider="heartbeat"

  operations

   op name="monitor" interval="5s" timeout="20s" prereq="nothing" id="stonithclone-op01"/

   op name="start" timeout="20s" prereq="nothing" id="stonithclone-op02"/

  /operations

 instance\_attributes id="stonithclone"

  attributes

   nvpair id="stonithclone01" name="hostlist" value="node1,node2"/

  /attributes

 /instance\_attributes

 /primitive

/clone

Load with the following command

sudo cibadmin -C -o resources -x /root/cluster/stonith.xml

\subsubsection*{Xen the cluster resources}
	
Now we can add the virtual machine XEN in the cluster.

Now we can add to the Xen virtual machine cluster resource. Lets say that we have a Xen to view the machine called vm01. The cofiguração and image files to keep us in vm01 /drbd0/xen/vm01/ in vm01.cfg and vm01-disk0.img respectively.

Edit /root/cluster/vm01.xml 

resources

 primitive id="vm01" class="ocf" type="Xen" provider="heartbeat"

  operations

   op id="vm01-op01" name="monitor" interval="10s" timeout="60s" prereq="nothing"/

   op id="vm01-op02" name="start" timeout="60s" start\_delay="0"/

   op id="vm01-op03" name="stop" timeout="300s"/

 /operations

 instance\_attributes id="vm01"

  attributes

   nvpair id="vm01-attr01" name="xmfile" value="/drbd0/xen/vm01/vm01.cfg"/

   nvpair id="vm01-attr02" name="target\_role" value="started"/

  /attributes

 /instance\_attributes

 meta\_attributes id="vm01-meta01"

  attributes

   nvpair id="vm01-meta-attr01" name="allow\_migrate" value="true"/

  /attributes

 /meta\_attributes

 /primitive

/resources

Load this file with the following command.

sudo cibadmin -C -o resources -x /root/cluster/vm01.xml

\subsubsection*{Tracking tools}

With the command "crm \_mon" you can track the inclusion of resources and in the cluster.

sudo crm\_mon Refresh in 14s...

The result of this command will be:

============

Last updated: Fri Jan 25 17:26:10 2008

Current DC: node2 (83972cf7-0b56-4299-8e42-69b3411377a7)

2 Nodes configured.

6 Resources configured.

============

Node: node2 (83972cf7-0b56-4299-8e42-69b3411377a7): online

Node: node1 (6bfd2aa7-b132-4104-913c-c34ef03a4dba): online

Clone Set: stonithclone

        stonithclone:0      (stonith:external/ssh): Started node1

        stonithclone:1      (stonith:external/ssh): Started node2

   vm01    (heartbeat::ocf:Xen):   Started node2

There is also a GUI available (graphical tool). To use it just set a password for the user "hacluster" with the following command and run the command "hb \_gui"

sudo passwd hacluster

password

re type password

sudo hb\_gui \&

\section{ANALYSIS OF RESULTS}

The fact work with LVM facilitated the mirroring of machines but the total security was not achieved because when we have to save the records in the course mirroring lost the last record and the bank needed to make the rollback. But the resumption of service in the event of the fall of the main server was made in seconds not creating inconvenience to users of the network.

Although the performance was higher with the use of XEN no details in this article because this item is not the purpose of it.

\section{CONCLUSION}

Looking up the sequence of servers religamento of the structure is made entirely stable and secure even in tests in Part 1 where only in processing memory was being implemented in PHP in the second machine (Server 2), after the fall continued smoothly.

In practical terms only at the end of the business can religar the main server (Server 1) again because of the time synchronization between the two high and for implying in the network for several minutes.

\end{document}